\documentclass[journal ]{new-aiaa}
\usepackage[utf8]{inputenc}
\usepackage{textcomp}

\usepackage{graphicx}
\usepackage{amsmath}
\usepackage{siunitx}
\usepackage{longtable,tabularx}
\usepackage{comment}
\newcommand{\CPS}{C_{P_{S}}}
\newcommand{\CPmin}{C_{P_{min}}}

\makeatletter
\def\@email#1#2{%
 \endgroup
 \patchcmd{\titleblock@produce}
  {\frontmatter@RRAPformat}
  {\frontmatter@RRAPformat{\produce@RRAP{*#1\href{mailto:#2}{#2}}}\frontmatter@RRAPformat}
  {}{}
}%
\makeatother

\title{Tracking stall cell dynamics at high Reynolds numbers}

\author{Badoui Hanna \footnote{PhD student/badoui.hanna@cstb.fr}}
\affil{CSTB,11 Rue Henri Picherit, Nantes, 44300, France}
\author{Bérengère Podvin\footnote{CNRS researcher/berengere.podvin@centralesupelec.fr}}
\affil{Laboratoire EM2C, CNRS, Centralesupélec, Université Paris-Saclay, 8-10 rue Joliot Curie, 91190 Gif/Yvette}
\author{Caroline Braud\footnote{CNRS researcher/caroline.braud@ec-nantes.fr}}
\affil{Nantes University, Centrale Nantes, LHEEA UMR 6598, 1, rue de la Noë,
44300, Nantes, FRANCE}

\begin{document}

\maketitle

\title{Tracking the dynamics of a stall cell at high Reynolds number}

\begin{abstract}
The spanwise organization of the flow over a thick airfoil  is investigated
using surface pressure measurements for a range of angles of attack around maximum lift and high Reynolds numbers  $O(10^6)$.
Locally strong  pressure fluctuations, which are not detected in the global lift coefficient, are shown to be associated with the presence of a stall cell.
The  stall cell  width is of the order of the chord length and increases linearly with the angle of attack, with a weak dependence on  the Reynolds number $Re_c$.
Its  dynamics  at Reynolds numbers larger than $10^6$ is dominated by a coherent motion in the spanwise direction with a characteristic velocity of order 0.1 U.
The motion can be decomposed into a  large-scale, low-frequency sweep with a Strouhal 
number  $St \sim 0.001$ combined with faster, smaller-scale oscillations.
The coherence of the stall cell makes it possible to track
global dynamics from local measurements.
\end{abstract}

\section{Introduction}

Understanding flow behavior in stall conditions is of utmost importance for
aerodynamic applications, including aeronautics, turbomachinery, and  wind turbines.
According to \cite{mccullough51}, three different types of stall can be identified, depending on the 
airfoil geometry. 
These include leading-edge stall for thin airfoils, combined trailing-edge leading edge stall,
and trailing edge stalls on thick airfoils.
The onset of stall is characterized by complex phenomena such as 
hysteresis and three-dimensional effects.
In particular, \cite{moss68} and \cite{gregory71} were among the first to provide experimental evidence of a cellular organization of the flow in the spanwise direction. 
\cite{winkelman80} used oil flow visualization  to identify mushroom-like or owl-faced structures consisting of counter-swirl patterns on 2D geometries that could not be attributed to end effects. 
\cite{broerenbragg01} showed that the appearance of stall cells was associated with trailing edge type stalls.

\cite{yonkatz98} showed that the number of stall cells was found  to increase with the wing  aspect ratio. 
Non-integer cell counts were reported and interpreted as evidence of unsteady switching between adjacent integer cell configurations
\cite{tairacolonius09} carried out a computational study of a flat plate at low Reynolds number at various angles of attack and found that the  spanwise aspect ratio of the plate needed to be large enough to observe stall cells.
The number of stall cells was also found to increase with the plate aspect ratio.
\cite{manni16} have numerically investigated the formation of stall cells 
 over a NACA 0012 airfoil of a high aspect ratio (10) at high Reynolds number ($10^6$). They found the presence of stall cell structures over a small range of angle of attack (within $2^\circ$), with a spacing of 1.4 to 1.8 chord lengths.
\cite{manolesos14} determined that the size of the stall cells increases with the angle of attack and with the Reynolds number.

Several models have been proposed for the origin of stall cells.
\cite{weihskatz83} proposed a mechanism in which the three-dimensional structures arose from a Crow-type instability
\cite{crow70} that would deform  the two-dimensional separation line.
\cite{spalart14}  used Prandtl's lifting line theory to derive a model that linked the appearance
of stall cells with a negative slope in the lift coefficient with respect to the angle of attack.
\cite{gross15} reached a similar conclusion and derived a formula for the spanwise cell spacing related to that slope.
Global stability analysis offers additional insight into the stall cell formation mechanism.
\cite{rodriguez11} provided a theoretical basis for the formation of stall cells using linear stability theory, according to which  
a stationary three-dimensional global mode consisting of two counter-rotating foci was identified at low Reynolds numbers.  
Further analyses have been carried out at higher Reynolds numbers combining linear stability analysis
with  URANS simulations.  \cite{plante21} provided evidence of a non-oscillatory stall  global mode for unswept wings. 
\cite{busquet21} used a similar approach to compute the  bifurcation scenario for  another 2D airfoil (OA209) and identified a low-frequency oscillation associated with the stall mode.

These theoretical results have been supported by  experimental observations.
\cite{zaman89} found evidence of a low-frequency oscillation in a range
of static stall around maximum lift regime with a 
Strouhal number based on projected height of the airfoil section and incoming velocity of 0.02.
This low frequency oscillation occurred in a narrow region  in the early stages of static stall (shallow stall) and was associated with a quasi-periodic switching of the flow between stalled and unstalled conditions.
\cite{broerenbragg01} associated the low-frequency instability with thin airfoils.
However, \cite{yonkatz98} showed that stall cell patterns on thick airfoils were responsible for large-amplitude fluctuations occurring at a very low frequency, which was significantly lower than that associated with bluff body vortex shedding.
Recently, \cite{neunaber22} observed such switches on a 2D thick airfoil at high Reynolds 
numbers, which they connected with spatial bistability in space between two chords on each side of the 
airfoil \cite{kn:braud24}. The bistability phenomenon was further characterized
in \cite{kn:braud24}, but the link between bistability and a stall cell 
 could not be confirmed due to the lack of extensive spanwise measurements.

This missing link is provided in the present study, which describes  
an experimental investigation of the same thick 2D airfoil at
high Reynolds numbers based on time-resolved pressure measurements along chords and transverse 
lines. The connection between global and local pressure fluctuations  is
examined in detail, allowing identification of a stall cell and tracking of its dynamics.
Special attention is given to the highest Reynolds number $Re=3.4$ $10^6$ which is the closest to wind energy applications.
The paper is organized as follows: after a presentation of the experiment, an analysis of pressure statistics is carried out.  The main characteristics of the stall cell are then extracted from local measurements. Complementary insight is provided by the application of Proper Orthogonal Decomposition.  

\section{Set-up}

\begin{figure}[htbp]
\centering
\includegraphics[width=0.9\textwidth]{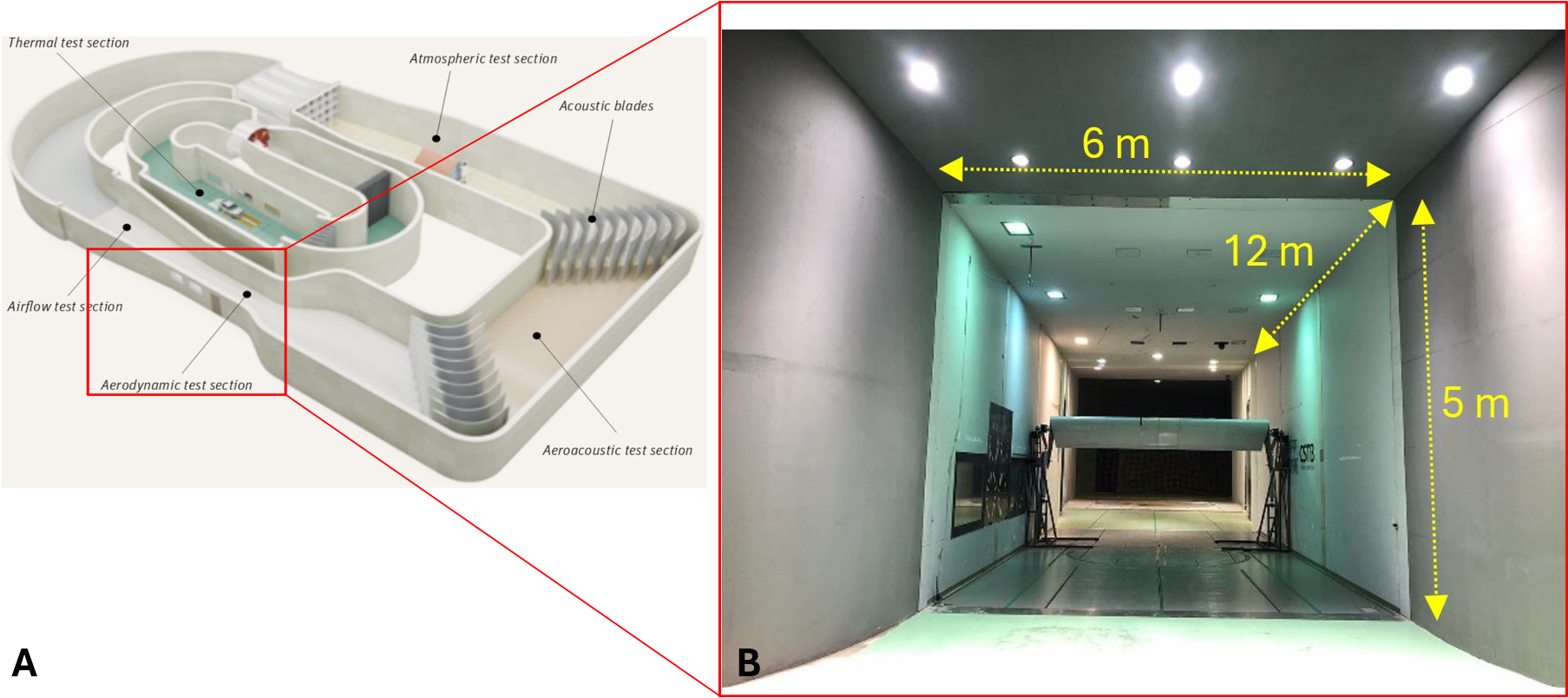}  
\caption{Aerodynamic test section of the CSTB wind tunnel.}
\label{windtunnel-setup}
\end{figure}

The experimental 2D geometry corresponds to the extruded section of a 2MW
commercial wind turbine operated by VALEMO at the Saint-Hilaire de Chaléons site in France. More details on the scan procedure is available in Braud {\it et al}  \cite{braud2025} and the 3D blade shape is available on the AERIS platform \cite{braud2024b}.
The location of the section is at 80\% of the rotor radius.
The airfoil is 5 m wide and has a full scale chord length of $c = 1.25 m$.
Test campaigns were conducted  
in the aerodynamic test section of the CSTB climatic wind tunnel
 at Nantes-France (see figure \ref{windtunnel-setup}) as part of the ANR/SNF MISTERY project.
The dimensions of the tunnel are respectively 6 m, 5 m and 12 m for width, height and length.
The airfoil has a maximum thickness of 20\%, located at 33\% of the
 chord, and a  maximum camber of 4\%, located approximately at 49\% of the chord.
Its aspect ratio is 4 and the maximum blockage ratio is 8\% when the angle of attack is  $24^\circ$.
   
\begin{figure}[htbp]
\centering
\includegraphics[width=0.9\textwidth]{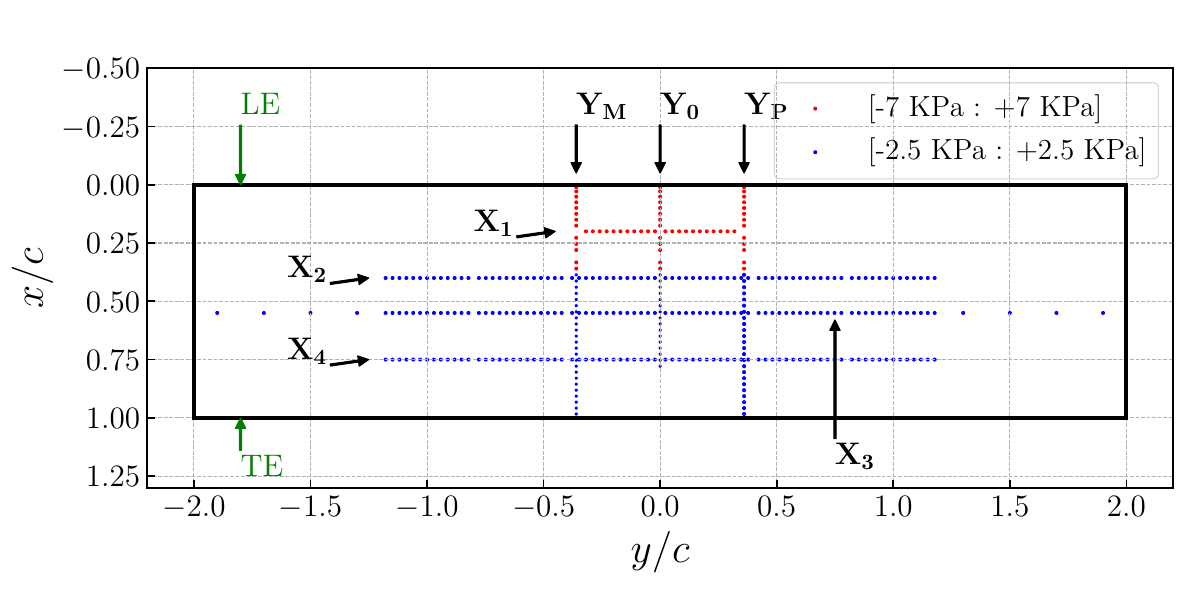}  
\caption{Pressure tap positions - top view.}
\label{Pressure_tap_top}
\end{figure}

The airfoil is equipped with a total of $N=476$ pressure taps, located along full chords as well
as on transverse lines on the suction side of the airfoil (see figure \ref{Pressure_tap_top}).
 The pressure taps were connected to five multiplexed EPS pressure scanners of 32 channels each, using
 $1.5 m$ long vinyl tubes with an internal diameter of $0.8 mm$. Two pressure sensor ranges were used
 depending on their location, 0 to 7 $kPa$ near the leading edge suction
 peak and 0 to 2.5 kPa elsewhere. The  precision  over the full measurement range was estimated to be $\pm 0.03 \%$. The transfer function of the whole system (tubes plus sensor cavity) has been measured off-line at a sampling frequency of 1024Hz (for the methodology, see, e.g., Holmes \& Lewis \cite{holmes1986}  and Whitmore et al. \cite{withmore1996}). The cut-off frequency is approximately 200Hz. The signal acquisition is performed using two National Instrument acquisition boards linked by real-time system integration for synchronization purposes. During the measurements, the sampling frequency was 512 Hz. During post-processing, the inverse transfer function is applied to the raw signals before the analysis. 

\begin{figure}[htbp]
\centering
\includegraphics[width=1.\textwidth]{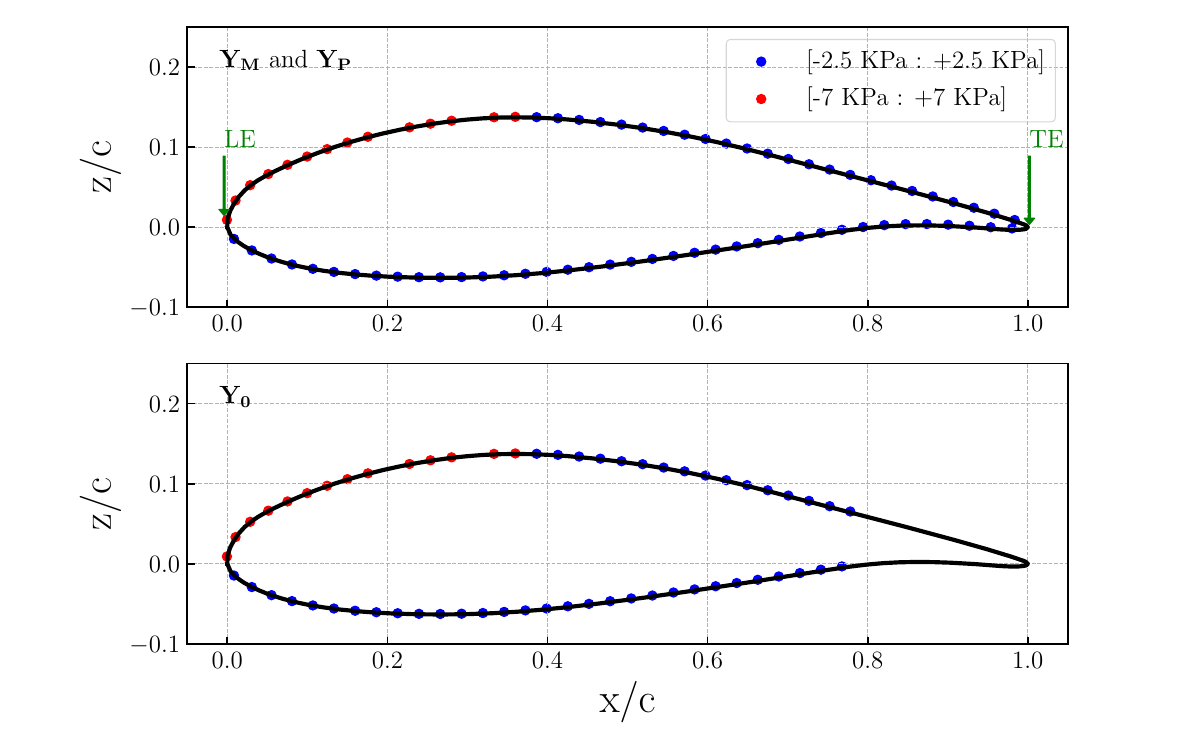}
\caption{Pressure tap positions along the chords.}
\label{Pressure_tap_crowns}
\end{figure}

The sensor locations presented in figure \ref{Pressure_tap_top} and \ref{Pressure_tap_crowns}  are distributed on three chords located at, $y = 0.36c, 0$ and $-0.36c$, denoted respectively as $Y_P$, $Y_0$ and $Y_M$, and  on four transverse lines located at $x = 0.2c, 0.4c, 0.55c$ and 0.75c  denoted respectively as $X_1$, $X_2$, $X_3$ and $X_4$.  The minimal spacing between the pressure taps is $0.026c$.
Four Reynolds numbers based on the incoming velocity and airfoil chord in the range $[5\times10^5$, $ 3.4\times10^6]$ were investigated.

\section{Pressure statistics}
\subsection{Global and local loads}

\begin{figure}[htbp]
\centering
\includegraphics[width=0.9\textwidth]{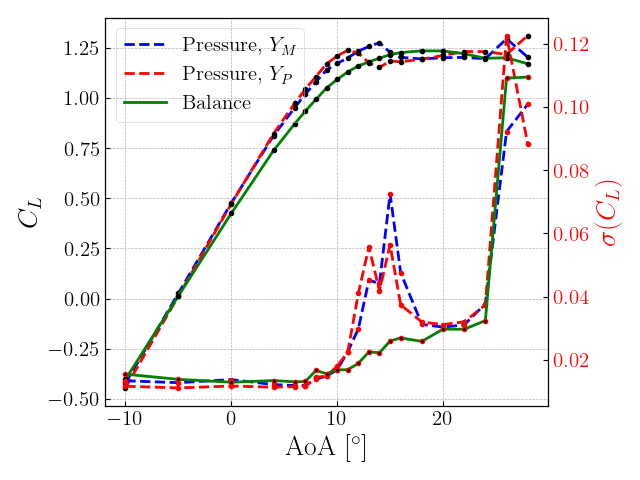}
\caption{Time-averaged (black points) and standard deviation (red points) of $C_L$  from the balance (green) and the integration of pressure taps around mid-chord (dash-blue and dash-red lines), $Y_M$ and $Y_P$ for $Re=3.4 \times 10^6$. }
\label{CL_comparison}
\end{figure}

Figure \ref{CL_comparison} compares the global average lift coefficient $<C_L>$ obtained
from balance measurements with the local lift coefficient obtained from pressure measurements at two crown locations ($Y_P$ and $Y_M$) on either side of mid-span for $Re_c=3.4\times10^6$.
The difference between global and local measurements of $⟨C_L⟩$ is strongest 
for $9^\circ \le AoA (^\circ) \le 12^\circ$ where a localized $<C_L >$ peak is  observed in the local measurements. 
This peak is followed by  a sharp negative slope over a narrow range of angles of attack, while the  global lift keeps increasing more slowly towards a maximum around  $AoA=18^\circ$ before decreasing slowly. This negative slope, which is also observed by Spalart \cite{spalart14}, is a first indication of the presence of stall cells.

The lift coefficient standard deviation is also very different for global and local measurements. Locally, two main fluctuation peaks are present:
\begin{itemize}
\item a wide peak for $AoA \in [12^\circ,15^\circ]$  at the inflection point of $⟨C_L⟩$, for which the negative slope of the mean lift is maximum,
 \item a higher one at high $AoA >24^\circ$ where stall occurs.
\end{itemize}
This second local peak is similar  to that observed in the global lift fluctuations.
In contrast, the level of the  first local peak is at least twice higher than the global lift fluctuation and only a steady increase is observed in the global measurements for $AoA \in [12^\circ,15^\circ]$. 
This indicates that the local, intense load fluctuations are not accounted for in global measurements
in that range. 
The pressure fluctuations in the local measurements were shown to increase  progressively in intensity with the Reynolds  number (not shown, see\cite{Hanna2024}), indicating that this local intensity increase is characteristic of high Reynolds numbers.

\subsection{Distribution along lines and chords}

\begin{figure}[htbp]
\centering
\includegraphics[width=0.9\textwidth]{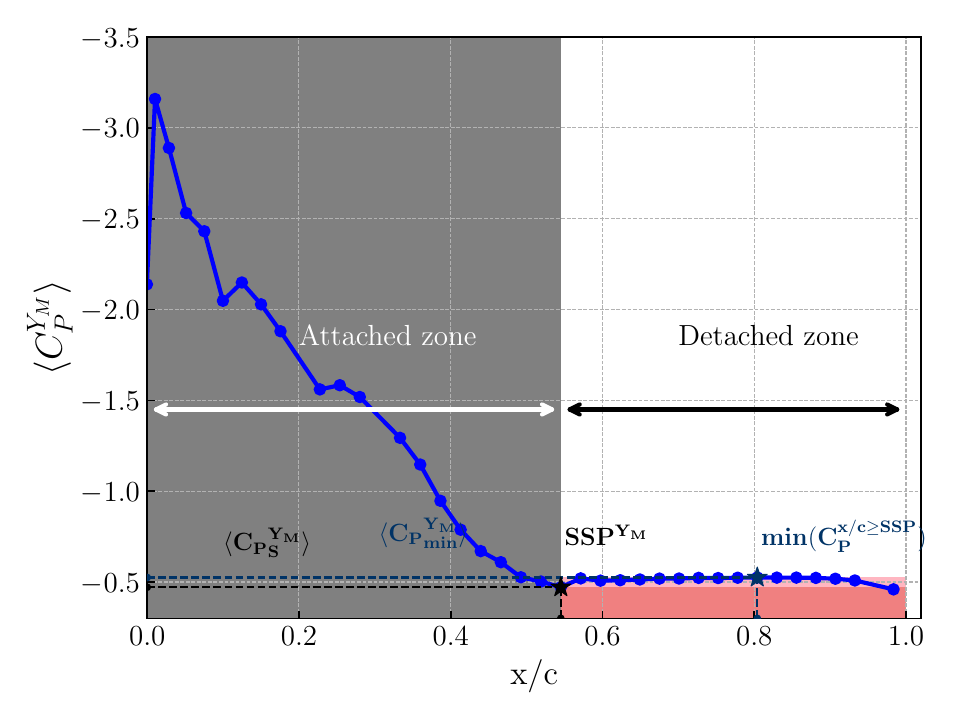}
\caption{Mean pressure distribution along the chord on the suction side of the airfoil $Y_M$ at $AoA=15^\circ$ and $Re=3.4 \times 10^6$. The colormap based on $\CPmin$ and 
$\CPS$ where is defined in the text.}
\label{Colorbar_crowns}
\end{figure}

This section focuses on the case $AoA=15^\circ$ and  $Re=3.4 \times $ $10^6$ that corresponds to the largest fluctuation level.  
The distribution of the average pressure coefficient along the chord $Y_M$ is shown in figure  \ref{Colorbar_crowns}. 
As discussed in \cite{kn:braud24}, it is characterized by a strong decrease followed by a plateau associated with separation, the origin of which corresponds to the steady separation point, denoted SSP and shown in figure
\ref{Colorbar_crowns}.
The steady separation point is typically defined as the first point where the pressure gradient becomes sufficiently close to zero (in practice, the criterion  $d<C_{p}>/dx < 1$ used in \cite{neunaber22} is applied).
Beyond the steady separation point, the pressure coefficient remains nearly constant and low in magnitude, and the flow is expected to be fully separated.
The pressure coefficient $C_p$ in the separation region  may take occasionally higher values  due to oscillations of the separated shear layer above the surface, but will remain close to the pressure coefficient 
 $\CPS$ measured at the SSP , with a higher bound $\CPmin \gtrapprox \CPS$. \\
This makes it possible to define a chord-based colormap for the pressure coefficient that allows
to distinguish attached and separated flow regions. 
This colormap is represented in figure \ref{Colorbar_crowns} and is defined as follows:
\begin{itemize}
\item If $C_p < \CPmin$ (grey): the flow is attached. 
\item If $C_p > \CPS$ (dark pink): the flow is fully separated. 
\item If $ \CPmin \le C_p \le \CPS $ (light pink): the flow is either attached (upstream of the SSP) or separated (downstream of the SSP). Due the strong pressure gradient upstream of separation, the size of the
upstream region is expected to be small and adjacent to the SSP. This intermittent region is actually located within the intermittent separation point peak and always ahead of SSP, as highlighted in \cite{kn:braud24}. However, because the Separation Point criteria can also be used instantaneously, it is preferred in the colormap representation. 
\end{itemize}

The chord-based colormap can be adapted to the full set of measurements by respectively replacing
$\CPmin$ and $\CPS$ with the minimum of $\CPmin$ and the maximum of $\CPS$ taken over the three chords $Y_i$ for $i \in \{ -,0, + \}$. These values will be denoted $\CPmin^{m}$ and $\CPS^{M}$.
Although perfect correspondence cannot be guaranteed, using this colormap makes it possible to determine regions where the flow is attached (above $\CPS^M$, in grey), fully separated (below  $\CPmin^m$, dark pink), and likely to be separated (between $\CPmin^m$ and $\CPS^M$, light pink).

\begin{figure}[htbp]
\centering
\includegraphics[width=0.9\textwidth]{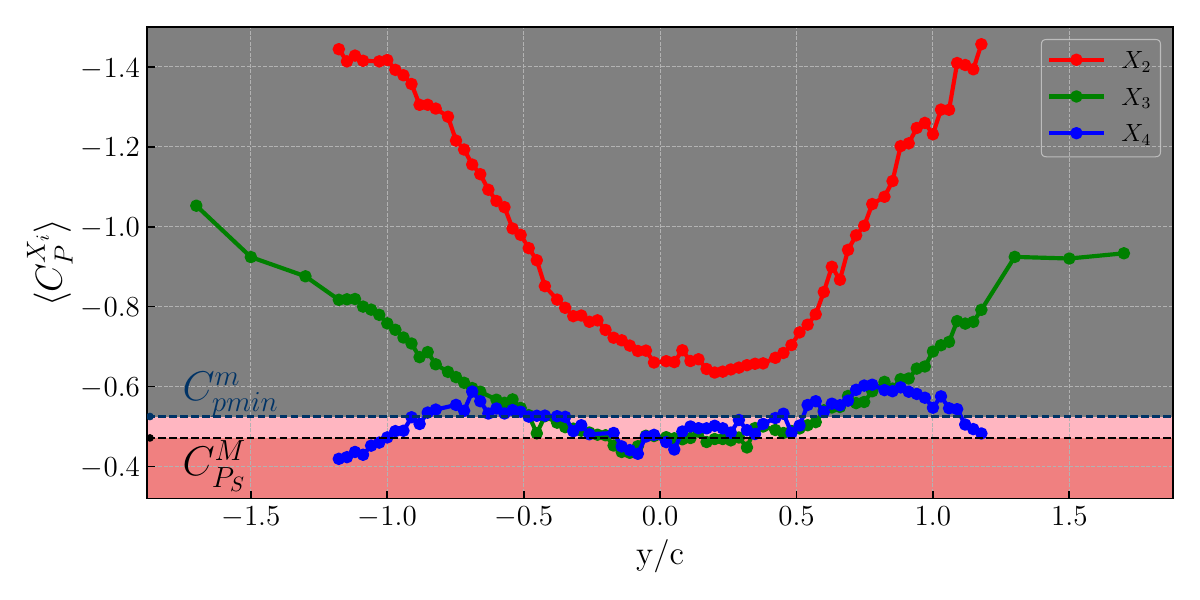}
\caption{Mean pressure distribution along the transverse lines at $AoA=15^\circ$ and $Re=3.4 \times 10^6$. The colormap based on $\CPmin^m=\mathrm{Min}_{Y_i} C_{pmin}^{Y_i}$ and 
$\CPS^{M}=\mathrm{Max}_{Y_i} \CPS^{Y_i}$ where $i \in \{-,0,+\}$ is defined in the text.}
\label{Colorbar_transverse}
\end{figure}

Figure \ref{Colorbar_transverse} shows  the time-averaged pressure coefficient for the different transverse lines using this colormap. 
It can be seen that the flow remains essentially attached at $X_2$.
Along the transverse line $X_2$, the average pressure reaches a maximum in the center region and decreases on each side. However the pressure levels remain low and the flow is expected to remain attached.
Similar variations are observed for $X_3$ but the pressure levels are higher, so that  the flow remains attached on the sides, while it is separated in the center region.
The pressure distribution along $X_4$ is similar to that of $X_3$ in the center region.
Moving away from the mid-span $y=0$, the pressure reaches  a local minimum close to the threshold value then increases again on the sides of the airfoil, so that unlike along $X_3$, the flow remains almost fully separated along $X_4$. 
This spanwise organization of the  pressure fluctuations is similar to that observed  by \cite{plante21} or \cite{mishraPhD} and is characteristic of the presence of a stall cell. 

\begin{figure}[htbp]
\centering
\includegraphics[width=0.9\textwidth]{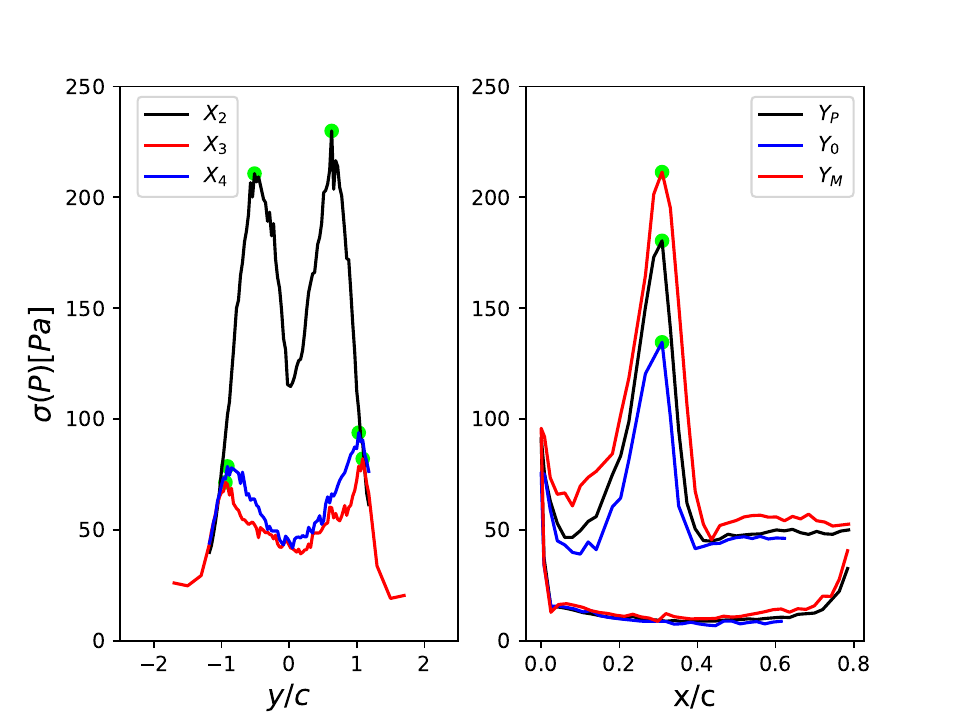}
\caption{Peaks of pressure standard deviation along lines (left) and chords (right), at $AoA=15^\circ$ and $Re=3.4 \times 10^6$. Green dots are maxima locations of ISP peaks.}
\label{SCFL-peaks}
\end{figure}

Figure \ref{SCFL-peaks} shows the pressure standard deviation along the lines and chords.
Figure \ref{SCFL-peaks}  left) shows that two maxima are present on each side of each transverse line
($y/c>0$ and $y/c<0$), in the range $y/c=\pm 0.5$ for $X_2=0.4c$,
and $y/c\sim\pm 1$ for $X_3=0.55c$ and $X_4=0.75c$. 
The presence of these two maxima indicates regions of high fluctuations, consistent with the intermittent presence of
separated flow. 
The maximum levels at $X_2$ are similar to those identified for the  chords $Y_M$ and $Y_P$ shown in  Figure \ref{SCFL-peaks}  right) - the lower levels for $Y_0$ are due to a smaller number of pressure taps, see figure \ref{Pressure_tap_crowns}).
The location of the fluctuation maximum, defined as the intermittent separation point by \cite{kn:braud24}, 
is around $x/c= 0.35$ along the three chords.
The location of the fluctuation maxima will be used below to describe the steady wall stall cell footprint.

\section{Stall cell characterization}

In this section we investigate the connection between local measurements and 
the stall cell over the range of angles of attack for which high fluctuation levels are observed. 
This corresponds to the first peak width of the lift fluctuation, $10^\circ<AoA<18^\circ$ (see figure \ref{CL_comparison}) and will be referred to as the stall cell regime. 

\subsection{Steady features}

\begin{figure}[htbp]
\centering
\begin{tabular}{c}
a)\includegraphics[width=1.\textwidth]{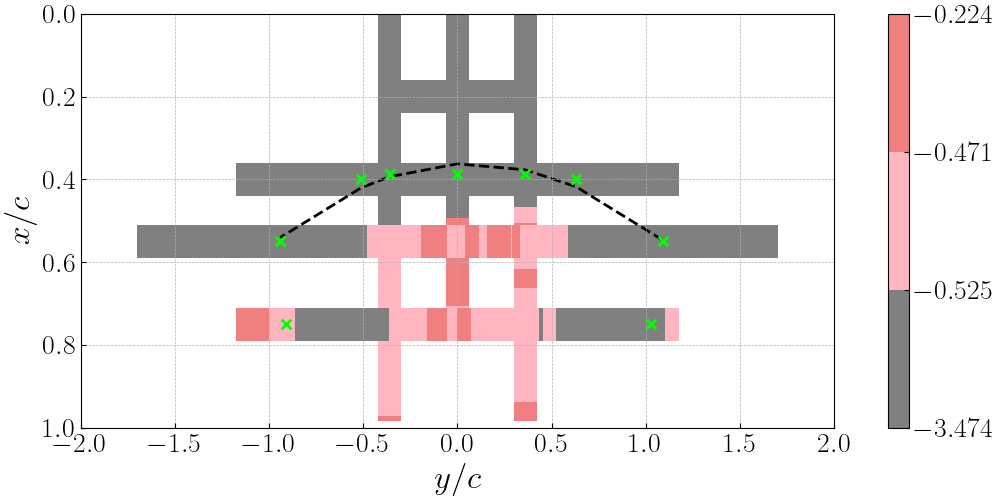}\\
b)\includegraphics[width=1.\textwidth]{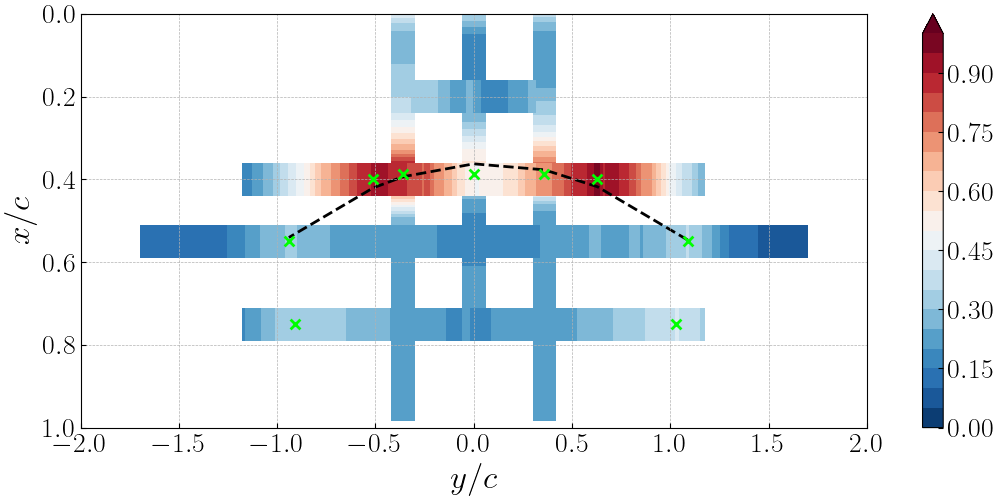}
\end{tabular}
\caption{Stall cell characterization at $AoA=15^\circ$ and $Re=3.4 \times 10^6$: a) Mean pressure coefficient, $<C_p>$, and b) normalized standard deviation of pressure, $\sigma(P)~/~max[\sigma(P)]$. The pink and gray colors of $<C_p>$ corresponds to the separation and attached flow region respectively. The black dashed line is obtained by joining the local maxima (green dots) along the three first lines and chords.}
\label{stallcell}
\end{figure}

 Pressure statistics such as  fluctuation maxima over the chords and transverse lines are
 first examined to determine the steady footprint of the stall cell. 
The time-averaged coefficient is represented  in Figure \ref{stallcell}a)  using the same colormap as in figure \ref{Colorbar_transverse}.
The locations of local pressure maxima along chords and lines  of figure \ref{SCFL-peaks} are reported in both plots as green dots.
The standard deviation map represented in Figure \ref{stallcell}b) shows that the pressure fluctuation levels reach a global maximum on $X_2$ in two regions centered around the chords $Y_P$ and $Y_M$.
An outline of the stall cell front was provided by a virtual line interpolating all fluctuation maxima  except  those of $X_4$ which are located at the rear of the cell. 
The interpolation was based on a second-order polynomial for
the highest two Reynolds numbers and on a linear fit for the lowest two Reynolds numbers, for reasons that will be detailed below.
The resulting intermittent separation line, represented on the plots as a black dashed line, delimits the front of the cell behind which the flow is separated. 
It will be referred to as the  stall cell separation line (SCSL).
The stall cell width can then be defined as the maximum spanwise distance between two points on the stall cell separation line.

\begin{figure}[htbp]
\centering
\includegraphics[width=0.9\textwidth]{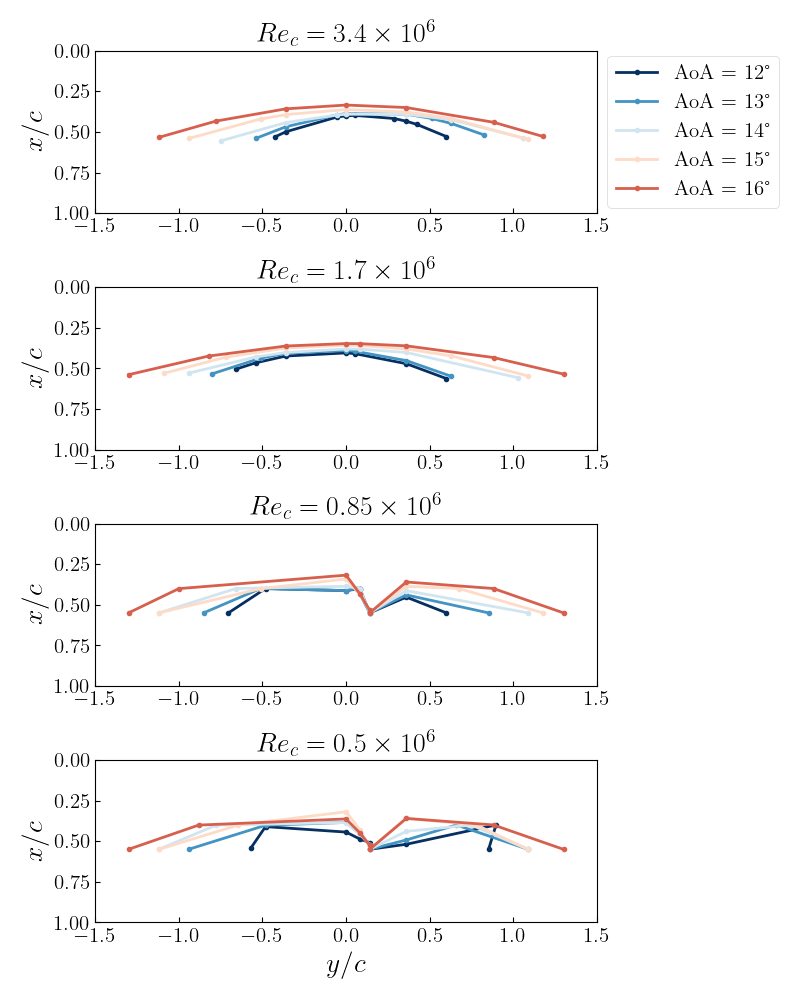}
\caption{Evolution of the stall cell front line with $AoA$ for different Reynolds numbers.}
\label{stallcellsize}
\end{figure}

\begin{figure}[htbp]
\centering
\includegraphics[width=0.7\textwidth]{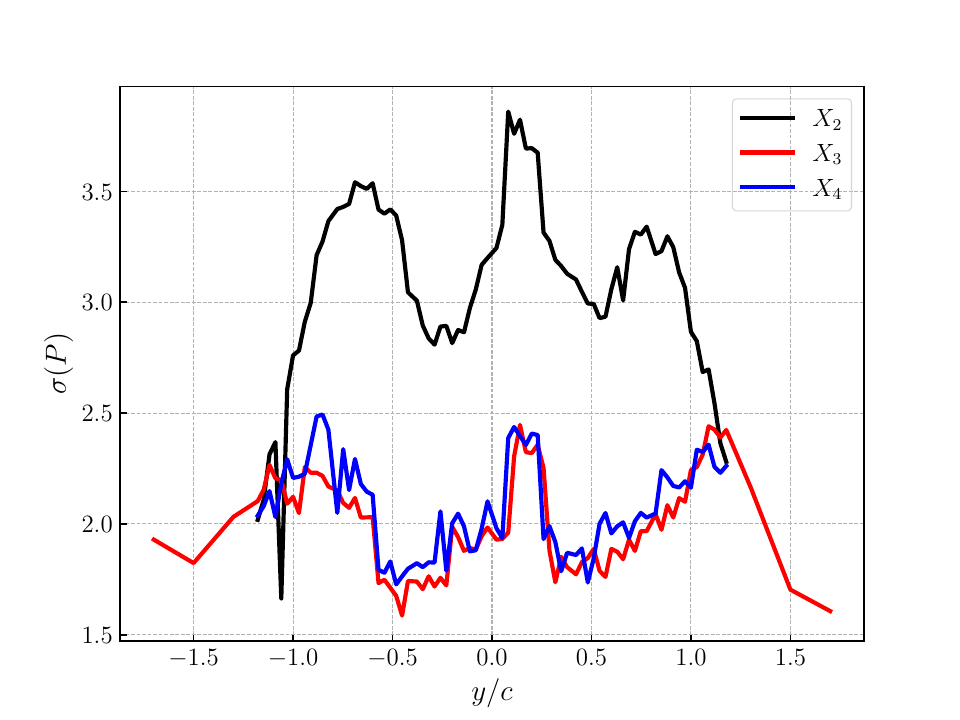}
\caption{Peaks of pressure standard deviation along lines at $AoA=15^\circ$ and $Re=0.5 \times 10^6$.}
\label{LowRe-stallcellsize}
\end{figure}

Figure \ref{stallcellsize} compares the separation lines for angles of attack in the stall cell regime  at different Reynolds numbers.
The stall cell width is on the order of $c$ at the onset of stall  ($AoA = 12^\circ$) and increases nearly linearly with the angle of attack at a given Reynolds number to reach more than $2c$ at $AoA = 16^\circ$.  
A small decrease of about 10-20\%  in the stall cell width can be observed as the Reynolds number increases from $1.7 \times 10^6$ to $3.4 \times  10^6$.
For the two lower Reynolds numbers $Re=0.5 \times 10^6$ and $Re=0.85 \times 10^6$, the location of the maximum on the central chord $Y_0$ was located on $X_3$ and not on $X_2$, unlike the higher Reynolds numbers. 
The presence of this kink motivated the use of linear interpolation and appeared to be consistent with a  splitting of the stall cell in two.
This  conjecture is supported
by examination of the pressure standard deviation along the transverse lines 
  for $Re_c=0.5\times 10^6$, shown in Figure \ref{LowRe-stallcellsize}. 
A third maximum at mid-span is observed for all lines, which is consistent with the presence of two stall cells. 
For the case $Re_c=0.85\times 10^6$, the maximum at mid-span was only observed at $X_2$, suggesting only a partial splitting (not shown). 
As stall cells  originate from the separated shear layer, this splitting may be explained by the decrease in shear intensity associated with lower Reynolds numbers. 

\subsection{Unsteady features }

\begin{figure}[htbp]
\centering
\includegraphics[width=1.\textwidth]{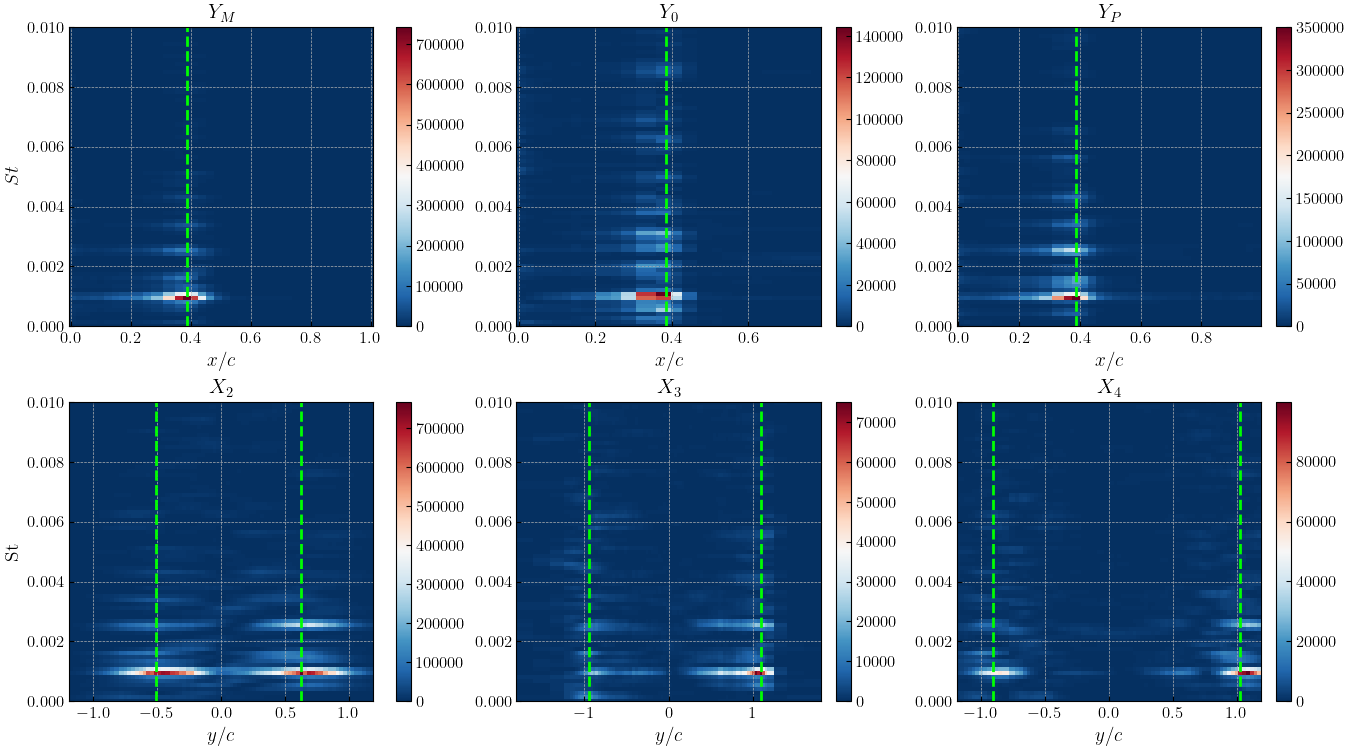}
\caption{Spectrogram of $C_p$ at $AoA=15^\circ$ and $Re=3.4 \times 10^6$. Green dotted lines corresponds to maxima of $\sigma(P)$ from dotted green points of figure \ref{SCFL-peaks}.}
\label{Uspec}
\end{figure}

In this section the time-dependent features of the stall cell are examined.
Figure \ref{Uspec} shows the spectrogram of $C_P$ for all lines and chords at $AoA=15^\circ$ and $Re=3.4 \times 10^6$. 
In all cases the local pressure fluctuations are characterized by maxima at low frequencies in the range below 1 Hz.
The frequencies $f$ appeared to slightly shift for different angles of attack and Reynolds numbers, but no coherent dependence on the angle of attack or the Reynolds number could be identified.
The frequencies are made non-dimensional by defining  Strouhal numbers $St$ 
as $St= f c \sin (AoA) / U$ where $c \sin (AoA)$ represents the projected chord area and 
and $U$ the incoming velocity. 
 Several local maxima are present in the range $St \in [10^{-3}, 10^{-2}]$ (this was also observed for other angles of attack and Reynolds numbers), although a clear global maximum at $AoA$=15$^\circ$ is observed at a very low frequency $St=0.001$.
Remarkably, the location of the spectral maxima lies on 
the stall cell separation line (local maxima locations are  represented with green lines on each plot), which suggests a strong coherence of the stall cell front.  

\begin{table}[h]
\centering
{\small
\begin{tabular}{|>{\centering\arraybackslash}p{1.2cm}|
                >{\centering\arraybackslash}p{2.65cm}|
                >{\centering\arraybackslash}p{2.65cm}|
                >{\centering\arraybackslash}p{2.65cm}|
                >{\centering\arraybackslash}p{2.65cm}|}
\hline
$AoA$ & $C(Y_P, Y_M)$ & $C(X_3^+, X_3^-)$ & $C(Y_P, X_3^+)$ & $C(Y_M, X_3^-)$ \\ \hline
13$^\circ$ & -0.793 & -0.285 & 0.644 & 0.583 \\ \hline
15$^\circ$ & -0.785 & -0.238 & 0.454 & 0.45 \\ \hline
16$^\circ$ & -0.551 & -0.31 & 0.223 & 0.241 \\ \hline
\end{tabular}
}
\caption{Correlation coefficients $C$ between signals taken at local fluctuation maxima  along the chords $Y_P$ and $Y_M$ and  the transverse line $X_3$ for $Re_c$ = $3.4\times10^6$ and AoA = 13$^\circ$, 15$^\circ$ and 16$^\circ$. The $\pm$ sign refers to the position of the maximum along $X_3$. See text for more details. }
\label{correl}
\end{table} 

To investigate this coherence in more detail, correlation coefficients between the pressure signals at 
the local fluctuation maxima are computed and reported in Table \ref{correl} for three angles of attack in the stall cell regime  at $Re=3.4$ $10^6$.
To alleviate notations,  $C(A, B)$ in the table represents the correlation coefficient between the fluctuation maximum locations 
(a.k.a. intermittent separation points) along A and B, where $A$ and $B$ refer to either lines or chords. A suffix $\pm$ is added 
to determine the maximum position $y>0$ or $y<0$ in the case of transverse lines. 

\begin{figure}[htbp]
\centering
\includegraphics[width=0.9\textwidth]{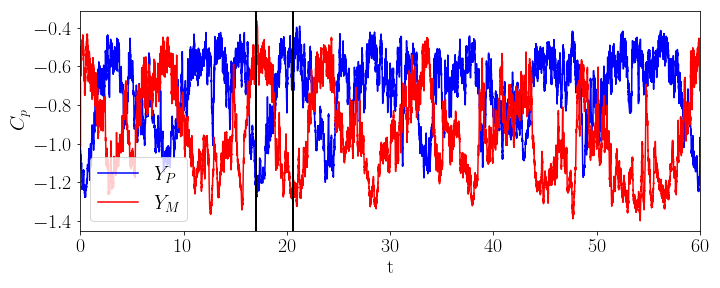} 
\caption{Evolution of the pressure coefficient measured at the intermittent separation point (maximum
fluctuation location) along $Y_P$ and $Y_M$
at $Re=3.4\times10^6$ and $AoA=15^\circ$. The vertical lines correspond
to the instants in figure \ref{cpinstant}.} 
\label{bistability}
\end{figure}

\begin{figure}[htbp]
\includegraphics[width=0.5\textwidth]{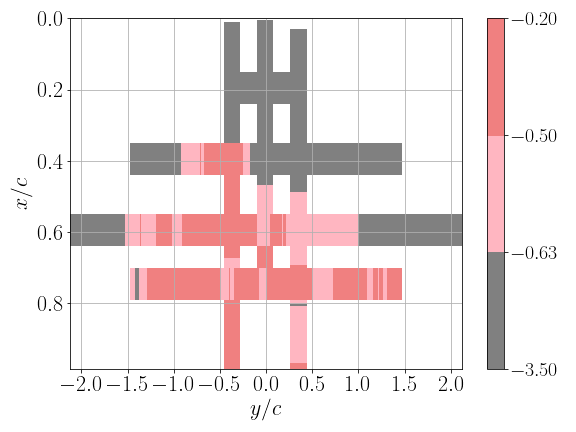} 
\includegraphics[width=0.5\textwidth]{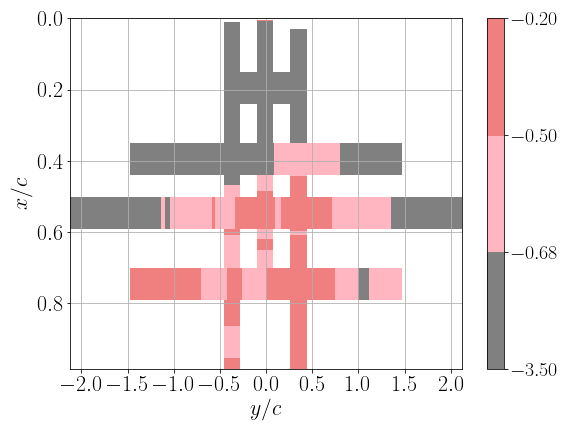} 
\caption{Pressure coefficient at instants $t=17.03 $ and $t= 20.60$ 
corresponding to  the vertical black lines in figure \ref{bistability}.
The dashed lines are obtained by interpolation of  the pressure fluctuation maxima
along the chords and transverse lines $X_2$ and $X_3$ as in figure \ref{stallcell}} 
\label{cpinstant}
\end{figure}

The correlation values are  strongest at $13^\circ$ and tend to decrease with the angle of attack. 
As observed in \cite{kn:braud24}, a strong anti-correlation is present  at the respective maxima along $Y_P$ and $Y_M$. 
The anticorrelation is still present but weaker for 
the two maxima along $X_3$, which are located closer to the trailing edge.
The same trends were reported  for $X_2$ and $X_4$ (not shown).
We note that generally higher correlations (in absolute value) were observed in the case of $X_2$. 
This difference could be  attributed to the presence of turbulent fluctuations associated with fully separated flow along $X_3$ and $X_4$, unlike along $X_2$.
Table \ref{correl} also shows that each of the maxima on $X_3$ is positively correlated with the intermittent separation point  on the chord located on the same side of the airfoil.
The correlation is maximal at zero time delay between all locations, which confirms 
the strong coherence of the stall cell dynamics.
This coherence is illustrated in 
figure \ref{bistability} which shows the evolution of $C_p$ at the intermittent
separation points along $Y_P$ and $Y_M$ for $AoA=15^\circ$
and $Re=3.4 \times 10^6$.
As detailed in \cite{kn:braud24}, the pressure coefficient is bimodal, and switches between a high-value and a low-value state, corresponding to attached and separated flow respectively.
Figure \ref{cpinstant} shows the instantaneous pressure coefficient map  at the two instants indicated in figure 
\ref{bistability} and associated with these two states. 
The colormap is similar to that used in figure \ref{Colorbar_transverse}, with the difference that instantaneous values are used  instead of time-averaged values for $\CPmin^m$ and $\CPS^M$. 
It can be seen that each state corresponds to an off-center position of the stall cell, and the pressure switches from one chord to the other. This corresponds to a global displacement of the  stall cell in the spanwise direction. 

\section{POD analysis}
\label{POD}

To investigate in more detail the global  motion of the stall cell, Proper Orthogonal Decomposition  (POD) \cite{kn:lumleyPOD} was applied to the fluctuating part of the pressure signal for each angle of attack and three Reynolds numbers ($0.8 \times 10^6$, $1.7 \times 10^6$, $3.4 \times 10^6$).
POD provides a decomposition of the spatial autocorrelation tensor into principal directions (aka spatial POD modes) so that the instantaneous fluctuating signal can be expressed as a combination of these modes.   
The pressure coefficient at the $j-th$ location and at time $t_k$ can thus be represented as
\begin{equation}
C_{p,j}(t_k) = <C_{p,j}> + \sum_{n=1}^{N_{POD}} a_n(t_k) \Phi_{n,j} 
\end{equation}
where $<C_{p,j}>$ is the pressure time-average value, $\Phi_{n,j}$ represents the value at position $j$ of the n-th spatial POD mode, and $a_n(t_k)$ is the amplitude of this mode at time $t_k$. 
$N_{POD}$ represents the dimension of the spatial autocorrelation and corresponds to the number of pressure taps considered. 
The modes $ \Phi_n$ are the eigenmodes  of the spatial
covariance of the pressure coefficient  and are associated with  eigenvalues $\lambda_n$, each of which represents the contribution of mode $n$ to the total variance.
They will be denoted as $ \Phi_n^G$ if POD is carried out over the full domain
$(N_{POD}=N)$, and
$ \Phi_n^L$ if POD is limited to a  line of measurements.
As the spatial modes are normalized, the eigenvalue $\lambda_n$ is 
also the variance of the mode amplitudes $a_n$.

\begin{figure}[htbp]
\centering
\includegraphics[width=0.9\textwidth]{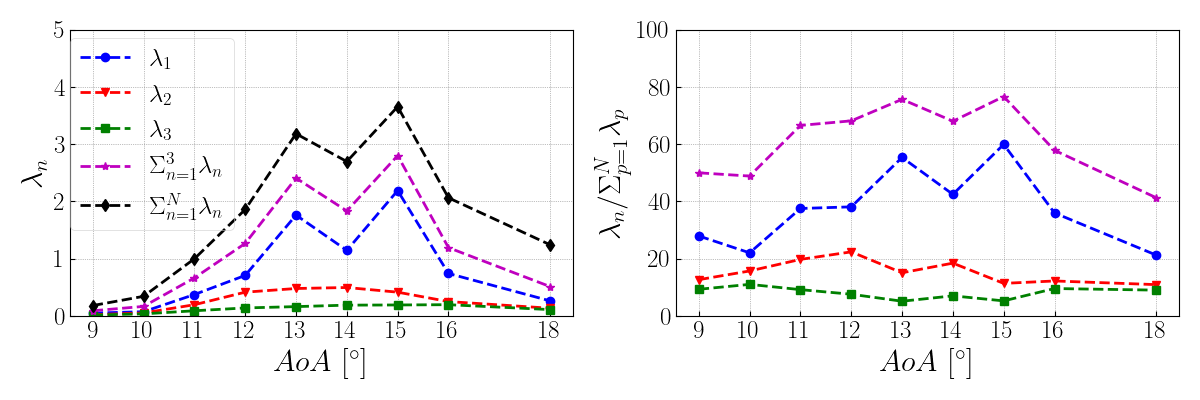}  
\caption{Global POD spectra at $Re=3.4 \times 10^6$ for the different angles of attack. a) Eigenvalues $\lambda_n$ ) Relative contribution of the eigenvalues. }
\label{lambda}
\end{figure}

The eigenvalue $\lambda_n$ of the  three most energetic modes  is represented in figure 
\ref{lambda} a) for all angles of attack at $Re=3.4 \times 10^6$ and their relative 
contribution reported in figure \ref{lambda} b).
The non-monotonic variations in the  total variance of the pressure observed in the stall cell regime region $[12^\circ,16^\circ]$ match those in the energy of the first mode. 
The second mode also reproduces these trends, but to a much  weaker extent, 
while the relative contribution of the third mode tends to decrease   in the region $[12^\circ,16^\circ]$.  
The relative contribution of the first two modes is largest  at theses angles of attack, with a value of about 60-70\%. 
This provides a quantitative measure of the strong spatio-temporal organization of the pressure fluctuations in the stall cell  regime.

\begin{figure}[htbp]
\begin{tabular}{ccc}
\includegraphics[width=0.28\textwidth]{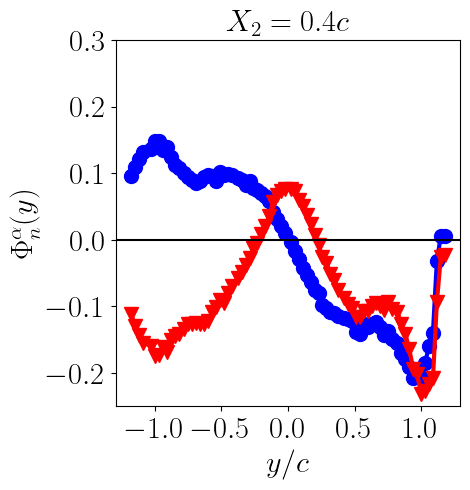} &
\includegraphics[width=0.38\textwidth]{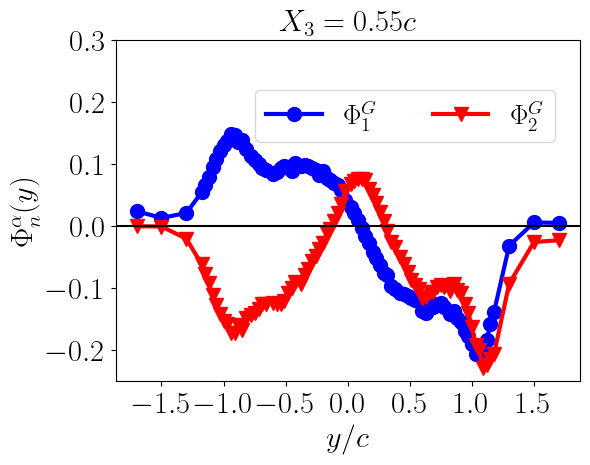} &
\includegraphics[width=0.28\textwidth]{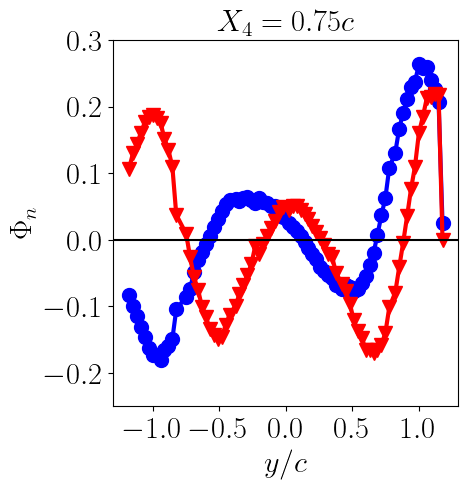} \\
\end{tabular}
\caption{POD spatial modes (defined over the full domain) at $AoA=15^\circ$ and $Re_c=3.4 \times 10^6$.}
\label{GlobalPhiAoA15}
\end{figure}

Figure \ref{GlobalPhiAoA15} shows the shape of the first two spatial POD modes at the angle of attack $15^\circ$  and $Re_c=3.4 \times 10^6$ along the transverse lines.
No significant changes in the shape of the modes were observed at other angles of attack (not shown),  except for the  change in the stall cell width evidenced in figure \ref{stallcellsize}.
The dominant mode  at $X_2$ and $X_3$ is antisymmetric and characterized by a sinusoidal shape with a $2c$ wavelength in the range $-1<y/c<1$, with extremal values located on the outer part of the domain at $y/c=\pm1$. 
Along $X_4$, this fundamental wavelength is still present (same period),  but in opposite phase with respect to $X_2$ 
and $X_3$, and modulated by its harmonic of wavelength $c$ ranging from $-0.5<y/c<0.5$.
Due to its antisymmetry, the spanwise average of the first mode along $X_2$ and $X_3$ is close to zero. This means that the first  POD mode makes almost no contribution to the global pressure coefficient, which explains the discrepancy between the local and global  pressure fluctuations evidenced in figure \ref{CL_comparison}.
 The second mode  also displays  a sinusoidal shape,
but is symmetric with respect to the mid-span. Its wavelength $c$ is half that  of the first mode for $X_2$ and $X_3$, with two extremal values at $y/c=\pm 1$, and one local extremum of opposite sign at  $y/c=0$.
A subharmonic modulation of wavelength $c/2$ is present at $X_4$.

\begin{figure}[htbp]
\centering
\includegraphics[width=0.8\textwidth]{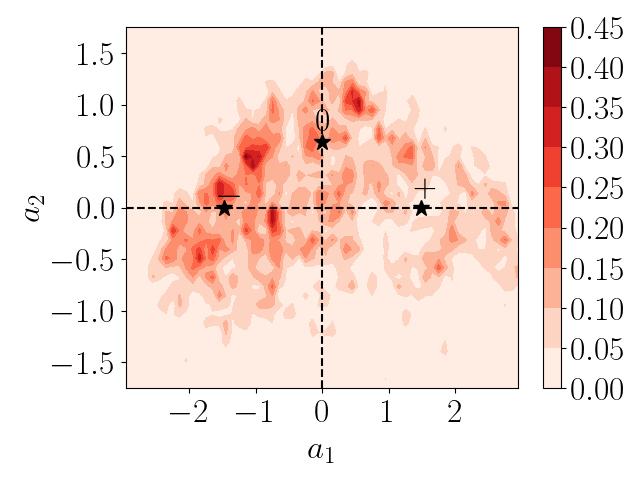} 
\caption{ Joint histogram of the POD amplitudes $a_1$ and $a_2$ 
at $AoA=15^\circ$ and $Re=3.4 \times 10^6$. The black stars correspond to the characteristic values used in figure \ref{reconstructionAoA15}. }
\label{ppa1a2AoA15}
\end{figure}

To describe the stall cell dynamics,
the joint histogram of the two dominant mode amplitudes $a_1$ and $a_2$ is represented 
in Figure \ref{ppa1a2AoA15}.
Similar histograms were observed for other angles of attack in the stall cell regime at
different Reynolds numbers. 
High values of $|a_1|$ are generally associated with small or negative values of $a_2$, corresponding to a region of high pressure
(separated flow) displaying a strong asymmetry on one side of the airfoil. 
Conversely, the highest values of $a_2$ are associated with values of
$|a_1|$ close to zero, consistent with a high pressure region located at mid-span.  
These observations are confirmed by  a significant 
correlation  coefficient of -0.4 between the absolute value of the amplitude $|a_1|$ and $a_2$.
This histogram also shows that some states (i.e. stall cell positions) are more visited than others, which suggests  complex, intermittent dynamics.

\begin{figure}[htbp]
\centering
\includegraphics[width=0.5\textwidth]{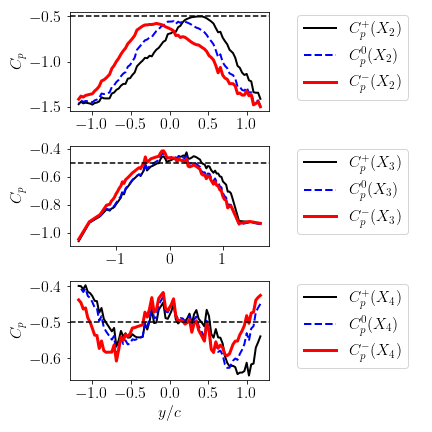} 
\caption{Comparison of reconstructed signals based on the first two POD modes 
at $AoA=15^\circ$ and $Re=3.4 \times 10^6$. }
\label{reconstructionAoA15}
\end{figure}

\begin{figure}[htbp]
\centering
\includegraphics[width=0.8\textwidth]{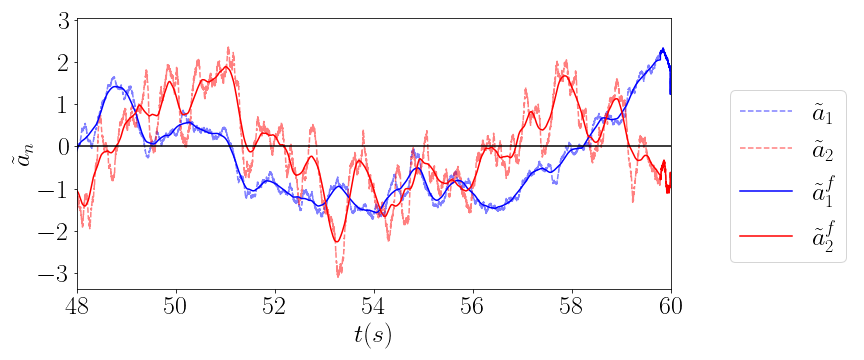} \\ 
\caption{ Evolution of the normalized POD amplitudes $\tilde{a}_1$ and $\tilde{a}_2$ (dashed lines) 
at $AoA=15^\circ$ and $Re=3.4 \times 10^6$. The solid lines correspond to  moving averages over 1 s. }
\label{a12}
\end{figure}

To investigate the respective action of the modes,
Figure \ref{reconstructionAoA15}  shows three  POD-based reconstructions of the pressure coefficient $C_p^\alpha$ on the  transverse lines $X_2$, $X_3$ and $X_4$.
The reconstructed pressure coefficients are defined as 
 $C_p^\alpha= a_1^\alpha \phi_1 + a_2^\alpha \phi_2$
for three states of $(a_1^\alpha,a_2^\alpha)$ where $\alpha \in \{-,0,+\}$.
These values correspond to
$(a_1^\pm,a_2^\pm)=(\pm \sqrt{\lambda_1},0)$ and 
$(a_1^0,a_2^0)=(0, \sqrt{\lambda_2})$ and are indicated  in Figure \ref{ppa1a2AoA15}. 
It can be seen that  the main difference between all three states is the spanwise position of the stall cell, although some  mild deformation of the cell is also observed. 
A switch of the stall cell  from $y<0$ to $y>0$ (from $C_p^-$ to $C_p^+$) corresponds to a change in the sign of the amplitude of the first mode (from positive to negative).
When the amplitude of the first mode reaches zero, the amplitude of the second mode
will typically be positive, which means that the stall cell is located at the middle of the airfoil ($C_p^0$). 
As the stall cell moves towards $y>0$, the amplitude $a_1$ becomes more negative
and $a_2$ decreases. 
An illustration of the   scenario  described above is shown  in figure \ref{a12}, which represents the time evolution of the first two normalized POD amplitudes $\tilde{a}_n=a_n/\sqrt{\lambda_n}$, $n=1,2$, corresponding to two back-to-back spanwise sweeps over a period of 12 $s$.

\begin{figure}[htbp]
\centering
\includegraphics[width=0.9\textwidth]{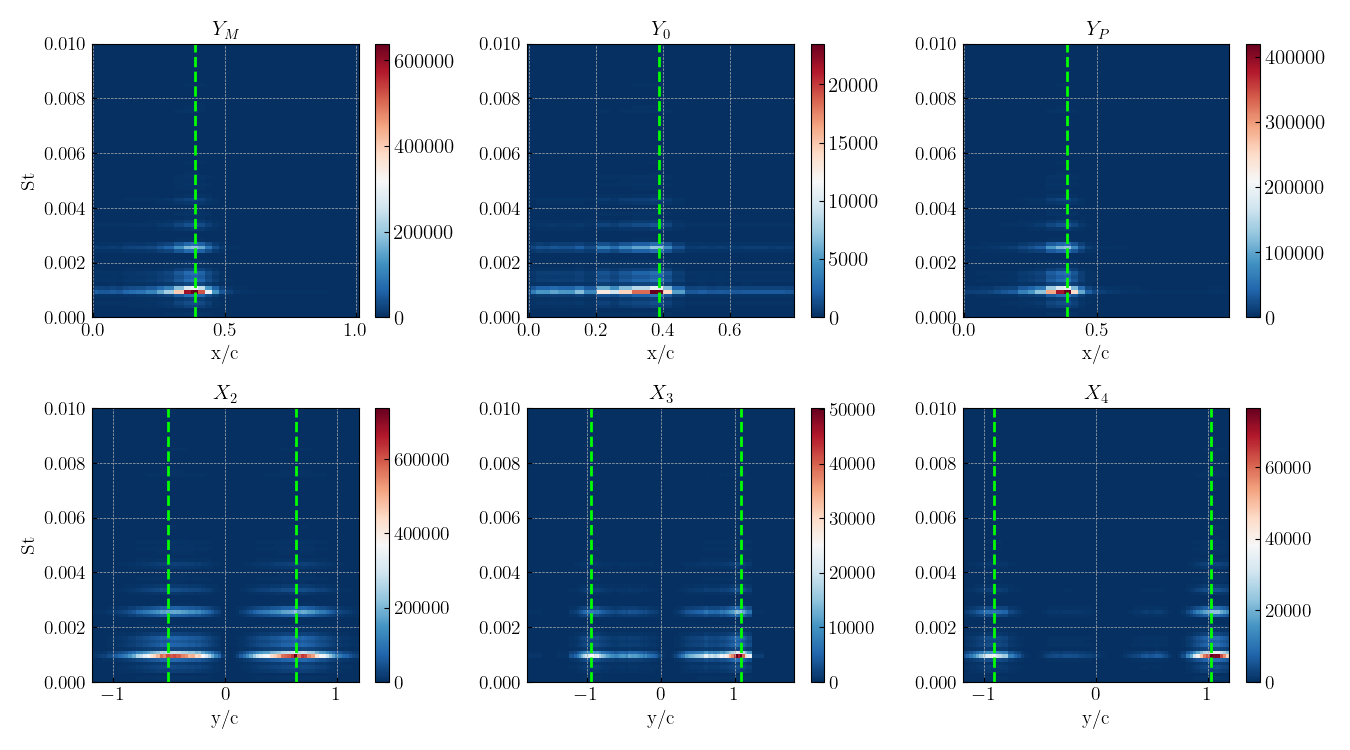}
\caption{Spectrogram of $C_p$ projected onto the first POD mode at $AoA=15^\circ$ and $Re=3.4 \times 10^6$. }
\label{Uspecproj}
\end{figure}

The spectrum of the reconstructed pressure coefficient limited to the first mode is shown in  figure \ref{Uspecproj}.
The good agreement  between figure \ref{Uspecproj}.and figure \ref{Uspec} shows  that the low-frequency dynamics, and in particular the maximum observed at $St=0.001$, are well captured by the first mode, which characterizes the large-scale motion of the stall cell in the spanwise direction. It can be noted that this frequency corresponds to the characteristic time over
which the stall cell moves from one side of the airfoil to the other (as seen 
in figure \ref{a12}).

\begin{figure}[htbp]
\begin{tabular}{ccc}
\includegraphics[width=0.28\textwidth]{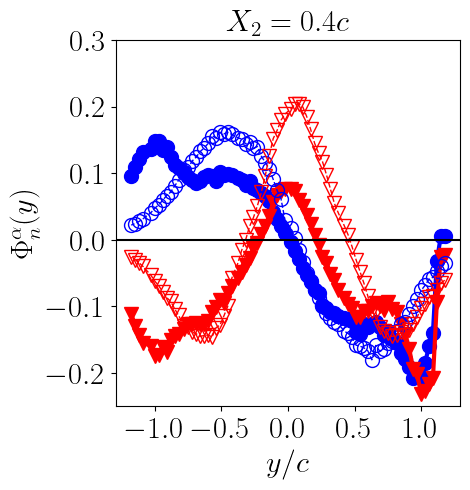} &
\includegraphics[width=0.38\textwidth]{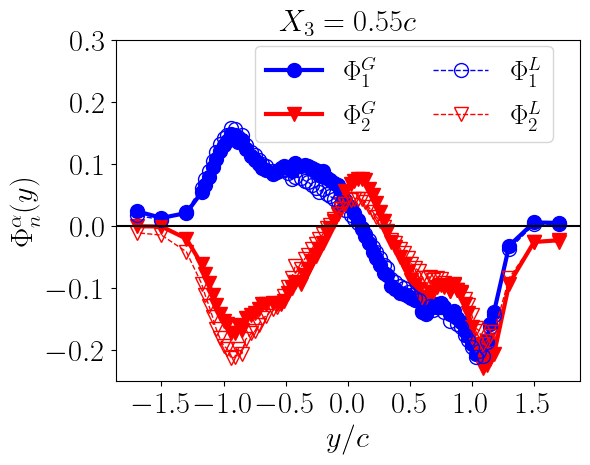} &
\includegraphics[width=0.28\textwidth]{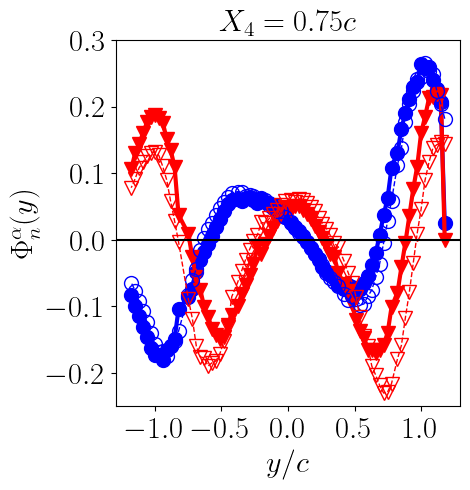} \\
\end{tabular}
\caption{Comparison of POD local modes with the local restriction of the global modes 
at $AoA=15^\circ$ and $Re_c=3.4 \times 10^6$. Full symbols: global POD modes; Open symbols: local POD modes.}
\label{GlobalLocalPhiAoA15}
\end{figure}

To determine whether these global flow dynamics can be tracked through limited measurements, 
Figure \ref{GlobalLocalPhiAoA15}  
compares the trace of the global modes $\Phi_n^G$  with local modes $\Phi_n^L$, 
which  were obtained by independent application of POD on individual transverse  lines.
It can be seen that the first two modes are largely similar for local and global decompositions on all transverse lines, particularly at $X_3$ (which corresponds to the position of the steady separation point for this angle of attack, as can be seen in figure \ref{Colorbar_crowns}). 
This confirms that the pressure fluctuations are globally coherent over the airfoil 
span. It also suggests that the dynamics of the stall cell  can be efficiently 
tracked through a single transverse line of measurements.
 
\begin{figure}[htbp]
\centering
\includegraphics[width=0.9\textwidth]{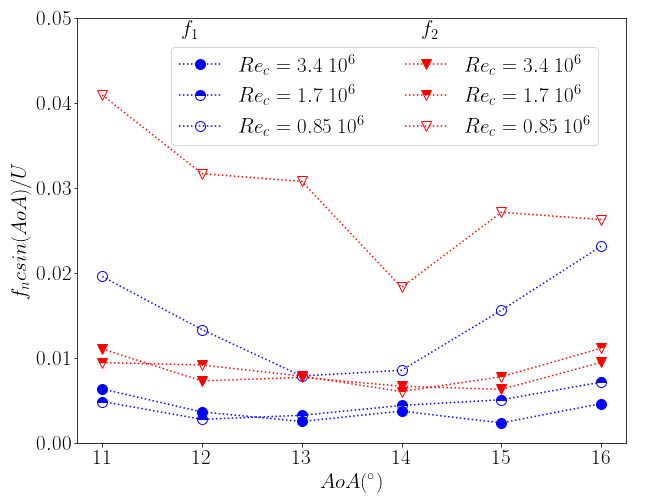}
\caption{Characteristic non-dimensional frequency $f_n$ of the dominant POD amplitudes $a_n$ for $n=1$ (blue symbols)
and $n=2$ (red symbols). }
\label{timescale}
\end{figure}

In all that follows we will consider local POD modes computed over the $X_2$ line, which corresponds
to the region of maximum fluctuations. 
For each mode $n$,  a characteristic frequency  $f_n $ for mode $n$ can be 
computed as 
\begin{equation}
f_n = \frac{\int f |\hat{a}_n(f) |^2 df}{\int |\hat{a}_n(f) |^2 df},
\end{equation}
where $\hat{a}_n(f)$ is the Fourier transform of the POD mode amplitude $a_n$.  
The nondimensional characteristic frequencies  for the first two dominant modes  
 are shown in figure \ref{timescale} 
at all angles of attack and three Reynolds numbers $\{0.8\times 10^6$, $1.7 \times 10^6$, $3.4 \times  10^6\}$.
It should be noted that the Strouhal numbers  based on $f_n$ are larger 
than that corresponding to the most energetic frequency since they integrate the effect of all frequencies.
A difference can be observed between the lowest Reynolds number and the two higher ones.
The characteristic frequencies are significantly higher at $Re=0.85 \times 10^6$ and show a local minimum in the stall cell regime.
In contrast, for the higher Reynolds numbers, the Strouhal numbers are relatively constant around respective values of $0.005$ and $0.01$   for the first and second POD modes.
  As a crude approximation, if we assume that $\phi_n(y) \approx e^{i k_n y} + C.C. $ (where $C.C.$ represents the complex conjugate) and
$a_n(t) \approx e^{- i f_n  t} + C.C.$, this defines a phase velocity for each mode $U_n=f_n/k_n$.
Since the frequency ratio  $f_2/f_1$ between the modes roughly corresponds to that of their spatial wavenumbers  $k_2/k_1$, this means that  the phase velocity of the first two modes is about the same, which corresponds to a global translation of the stall cell at speed $U_c$, as
$C_p(y,t)=\sum_n a_n(t) \phi_n(y) \approx f(y-U_c t)$.
An estimate for the global convection speed $U_c$ of  the stall cell 
can then be given by $U_c \sim 0.1 U$.
Results at Reynolds numbers larger than $10^6$ are therefore consistent with the superposition of a low-frequency sweep and faster, small-scale oscillations, as illustrated in figure \ref{a12}.

\section{Conclusion}

 Time-resolved pressure measurements were taken along the chords and span of a moderately thick 2D airfoil at high Reynolds numbers in the range $[0.5, 3.4] \times 10^6$.
Evidence of a stall cell is found in a range of angle of attacks $[12^\circ,16^\circ]$ which is close to the lift maximum  and where local loads exhibit a negative slope. 
The steady footprint of the stall cell can be outlined using a separation line based on local fluctuation pressure maxima. 
The stall cell width  defined from the separation line  is found to increase roughly linearly with the angle of attack.
 It tends to slightly decrease with the Reynolds number  for $Re_c>10^6$. At the two lowest Reynolds numbers $Re_c \le 10^6$,  the cell observed at the higher Reynolds numbers appears to split in two smaller cells.
Fluctuations on the separation line are dominated by a range of frequencies lower than 1 Hz and correspond to a motion that is predominantly in the spanwise direction.
The stall cell motion can be decomposed into 
 a  low-frequency, large-scale  sweep  with a Strouhal number
of order $O(10^{-3})$,
and  faster, smaller-scale oscillations.  It is characterized by a global  convection velocity of order $0.1 U$.

The results of the study have direct implications  for the applications: firstly, stall cell characteristics, such as the frequencies and fluctuation levels measured in the region directly upstream of separation, depend on the Reynolds number, which shows the importance of carrying out large-scale experiments.
Secondly,  the global dynamics of the stall cell can be tracked
from local measurements taken along a transverse line, which opens up interesting possibilities for flow estimation from sparse sensor data and real-time control. 
\section{Acknowledgements}

This research was performed within the French-Swiss project MISTERY funded by the French National
Research Agency (ANR PRCI grant no. 266157) and the Swiss National Science Foundation (grant no. 200021L 21271)

\bibliography{bib}

\end{document}